\documentclass[doublecol]{epl2}

\usepackage{times}
\usepackage{fmtcount} 
\usepackage{hyperref}
\hypersetup{
breaklinks=true,
colorlinks=true,
citecolor=blue,
linkcolor=blue,
filecolor=blue,
urlcolor=blue
}

\title{Valence-bond theory of highly disordered quantum antiferromagnets}

\author{S. Zhou \inst{1} \and Jos\'e A. Hoyos\inst{2}\thanks{E-mail: \email{jh146@phy.duke.edu}} \and V. Dobrosavljevi\'c\inst{1} \and E. Miranda\inst{3}}
\shortauthor{S. Zhou, J. A. Hoyos, V. Dobrosavljevi\'c and E. Miranda}

\institute{
	\inst{1} National High Magnetic Field Laboratory - Florida State University, Tallahassee, Florida 32310 \\
	\inst{2} Department of Physics - Duke University, Durham, North Carolina 27708 \\
	\inst{3} Instituto de F\'{\i}sica Gleb Wataghin - Unicamp, C.P. 6165, Campinas, S\~ao Paulo 13083-970, Brazil
}


\abstract{
We present a large-$N$ variational approach to describe the magnetism
of insulating doped semiconductors based on a disorder-generalization
of the resonating-valence-bond theory for quantum antiferromagnets.
This method captures all the qualitative and even quantitative predictions
of the strong-disorder renormalization group approach over the entire
experimentally relevant temperature range. Finally, by mapping the
problem on a hard-sphere fluid, we could provide an essentially exact
analytic solution without any adjustable parameters.
}

\begin{document}

\maketitle


The metal-insulator transition (MIT) in doped semiconductors (DS)
\cite{paalanen91} is one of the most fundamental, yet theoretically
less understood problems in condensed matter physics. Even aside from
their pivotal technological role, the DS have long been recognized
as a bellwether system for the study of quantum criticality at the
MIT. Careful transport experiments have revealed sharply defined critical
behavior, although with exponents inconsistent with early theoretical
predictions \cite{sarachik95}.

What are the basic physical processes that drive this transition and
localize the electrons? Important clues have been provided by the
thermodynamic response on the insulating side. Here, no magnetic ordering
has been experimentally observed down to the lowest temperatures,
while both the spin susceptibility and the specific heat display signatures
of randomly interacting localized magnetic moments \cite{paalanen91,paalanenetal88}.
This puzzling behavior was largely explained by the Bhatt-Lee (BL)
theory \cite{bhattlee82} of random singlet (RS) formation, using
a strong-disorder renormalization group (SDRG) approach \cite{madasguptahu}. 

The remarkable success of the BL theory provides strong support to
the early ideas of Mott \cite{mott-book90}, who first emphasized
that strong Coulomb repulsion may localize the electrons by converting
them into localized magnetic moments. According to this picture, the
MIT in DS should be viewed as a disordered version of the Mott transition,
a phenomenon dominated by strong correlation effects. An appropriate
theory should then be able to describe both the local moment magnetism
in the insulator and the transmutation of these local moments into
conduction electrons on the metallic side of the MIT. Unfortunately,
the SDRG approach of BL, which was so successful in the insulator,
is difficult to extend across the transition. 

The essential challenge, therefore, is to develop an alternative approach
to Mott localization in a strongly disordered situation, one that
\emph{at the very least} can reproduce the RS physics of Bhatt and
Lee. An attractive avenue to describe strong correlations has emerged
in the last twenty years from studies of various Mott systems, based
on resonating-valence bond (RVB) ideas of Anderson \cite{anderson87}
and others. At the mean field level, these theories provide variational
wavefunctions for quasiparticle states, which become exact in appropriate
large-$N$ limits \cite{affleck-marston88}. Very recent work has
extended similar variational studies to disordered systems, providing
a description of phenomena such as disorder-induced non-Fermi liquid
behavior \cite{NFLreview}, but did not address the physics of inter-site
spin correlations central to the BL paradigm. 

In this Letter we examine an appropriate $t$-$J$ model capable of
describing the Mott transition in a disordered environment. While
the large-$N$ limit of this model generally reduces to an RVB-like
variational problem, here we concentrate on the localized 
($t\rightarrow0$) limit in the presence of strong positional disorder modeling the insulating
DS. We show that: (i) the large-$N$ formulation quantitatively reproduces
all the key features of the RS regime; (ii) an accurate analytic solution
of the variational problem can be thus obtained, providing closed
form expressions for various physical quantities; and (iii) the approach
can be directly extended to the metallic side, eliminating the main
stumbling block in attacking the MIT in DS. 


\textit{Model and large-N formulation.} We start with the large-$N$
formulation of the two-orbital $t$-$J$ model, 
\begin{eqnarray}
\mathcal{H} \!\!\!\! & = & \!\!\!\! \sum_{\mathbf{k},\sigma}(\varepsilon_{\mathbf{k}}+\varepsilon_{o})c_{\mathbf{k}\sigma}^{\dagger}c_{\mathbf{k}\sigma}+\sum_{i\neq j,\sigma}t_{ij}\tilde{f}_{i\sigma}^{\dagger}\tilde{f}_{j\sigma}\label{htj}\\
 \!\!\!\!& & \!\!\!\! +\frac{1}{2N}\sum_{i\neq j}J_{ij}\mathbf{S}_{i}\cdot\mathbf{S}_{j}+\frac{V}{\sqrt{N}}\sum_{i,\mathbf{k},\sigma}(e^{i\mathbf{k}\cdot\mathbf{r}_{i}}c_{\mathbf{k}\sigma}^{\dagger}\tilde{f}_{i\sigma}+\mathrm{H.c.}),\nonumber 
\end{eqnarray}
under the constraint of no double occupancy on the $\tilde{f}$-orbital.
Here each lattice site corresponds to a donor or acceptor which is
randomly distributed in a periodic-boundary 3D cube of volume $V_{0}=N_{0}/\rho_{0}$,
where $N_{0}$ is the number of dopant sites and $\rho_{0}$ is the
doping concentration. We stay at half-filling for the uncompensated
DS, $\sum_{\mathbf{k},\sigma}c_{\mathbf{k}\sigma}^{\dagger}c_{\mathbf{k}\sigma}+\sum_{i\sigma}\tilde{f}_{i\sigma}^{\dagger}\tilde{f}_{i\sigma}=N_{0}N/2$.
The $c$-orbital represents the semiconductor conduction band with
dispersion $\varepsilon_{\mathbf{k}}$, lying at an energy $\varepsilon_{o}$
above the hydrogenic 1$s$ impurity bound state (the $\tilde{f}$-orbital),
and $V$ is the hybridization between them. $\mathbf{S}_{i}$ is the
SU($N$) spin operator of the $\tilde{f}_{i}$-orbital. The hopping
between the hydrogenic bound states \cite{andre81} falls off exponentially
with distance $r_{ij}=|\mathbf{r}_{i}-\mathbf{r}_{j}|$, 
$t_{ij}=t_{0}\exp(-r_{ij}/a)$ for $r_{ij}\gg a$,
the Bohr radius of the bound state.
Consequently, the antiferromagnetic super-exchange coupling 
\begin{equation}
J_{ij}=J_{0}\exp(-2r_{ij}/a),
\label{tjs}
\end{equation}
 where $J_{0}\sim t_{0}^{2}$ (see footnote \footnote{It is well known (and expected) that corrections to $J_{ij}$ exist due to
anisotropy and other effects. For instance, in $d=3$ there is an additional factor of $(r_{ij}/a)^{5/2}$ 
multiplying $J_{ij}$ when $r_{ij}\gg a$ \cite{andre81}.For our purposes, these
corrections can be safely neglected in face of the highly
disordered character of the dilute (insulating) regime. They
only provide subleading (logarithmic) corrections as we
confirmed numerically. Furthermore, this allows us to
directly compare our results with those of the Bhatt-Lee theory
 \cite{bhattlee82}, which also neglects them.}). 
 The projected Hilbert space of the $\tilde{f}$-orbital
can be treated in the slave-boson formalism $\tilde{f}_{i\sigma}^{\dagger}=b_{i}f_{i\sigma}^{\dagger}$
enslaved to a constraint on each site $\sum_{\sigma}f_{i\sigma}^{\dagger}f_{i\sigma}+b_{i}^{\dagger}b_{i}=N/2$.

In this Letter, we focus on the insulating side of the uncompensated
DS $\rho_{0}<\rho_{c}$ ($\rho_{c}^{1/3}a\approx0.25$ for Si:P) where
the average inter-site distance $\Lambda=\rho_{0}^{-1/3}\gg a$, which
implies that $t_{ij}\rightarrow0$. In this limit, the effective hybridization
$bV$ goes to zero as $b\rightarrow0$, and the electrons become Mott
localized on singly-occupied $\tilde{f}$-orbitals. This results in
an effective Heisenberg Hamiltonian for the insulating uncompensated
DS, $\mathcal{H}=\frac{1}{2N}\sum_{i\neq j}J_{ij}\mathbf{S}_{i}\cdot\mathbf{S}_{j}$.
The magnetic behavior of such a disordered Heisenberg system was largely
explained by Bhatt and Lee via the SDRG method. Here we investigate
the system within the large-$N$ theory \cite{affleck-marston88,jones89prb},
which leads to an effective mean-field Hamiltonian through the saddle-point
approximation, 
$$\mathcal{H}=-\frac{N}{16}\sum_{i\neq j}J_{ij}\left(\Delta_{ij}^{\ast}\hat{\Delta}_{ij}+\mathrm{H.c.}-|\Delta_{ij}|^{2}\right),$$
with the constraint $\sum_{\sigma}f_{i\sigma}^{\dagger}f_{i\sigma}=N/2$
(of self-conjugate spins) implemented through the local Lagrange multiplier
$\lambda_{i}$. Here, $\hat{\Delta}_{ij}=2\sum_{\sigma}f_{i\sigma}^{\dagger}f_{j\sigma}/N$
are valence bond (VB) operators and $\Delta_{ij}=\left\langle \hat{\Delta}_{ij}\right\rangle $
are variational parameters which minimize the free energy. They are
solved self-consistently at $N\rightarrow\infty,$ for a given sample
realization and temperature. The results are then averaged over 20
sample realizations. 

\addtocounter{footnote}{1}
\footnotetext{We have carefully verified that all our numerical results are robust with respect to finite-size effects.\label{foot2}}

\begin{figure}
\begin{centering}
\includegraphics[clip,width=0.85\columnwidth,keepaspectratio]{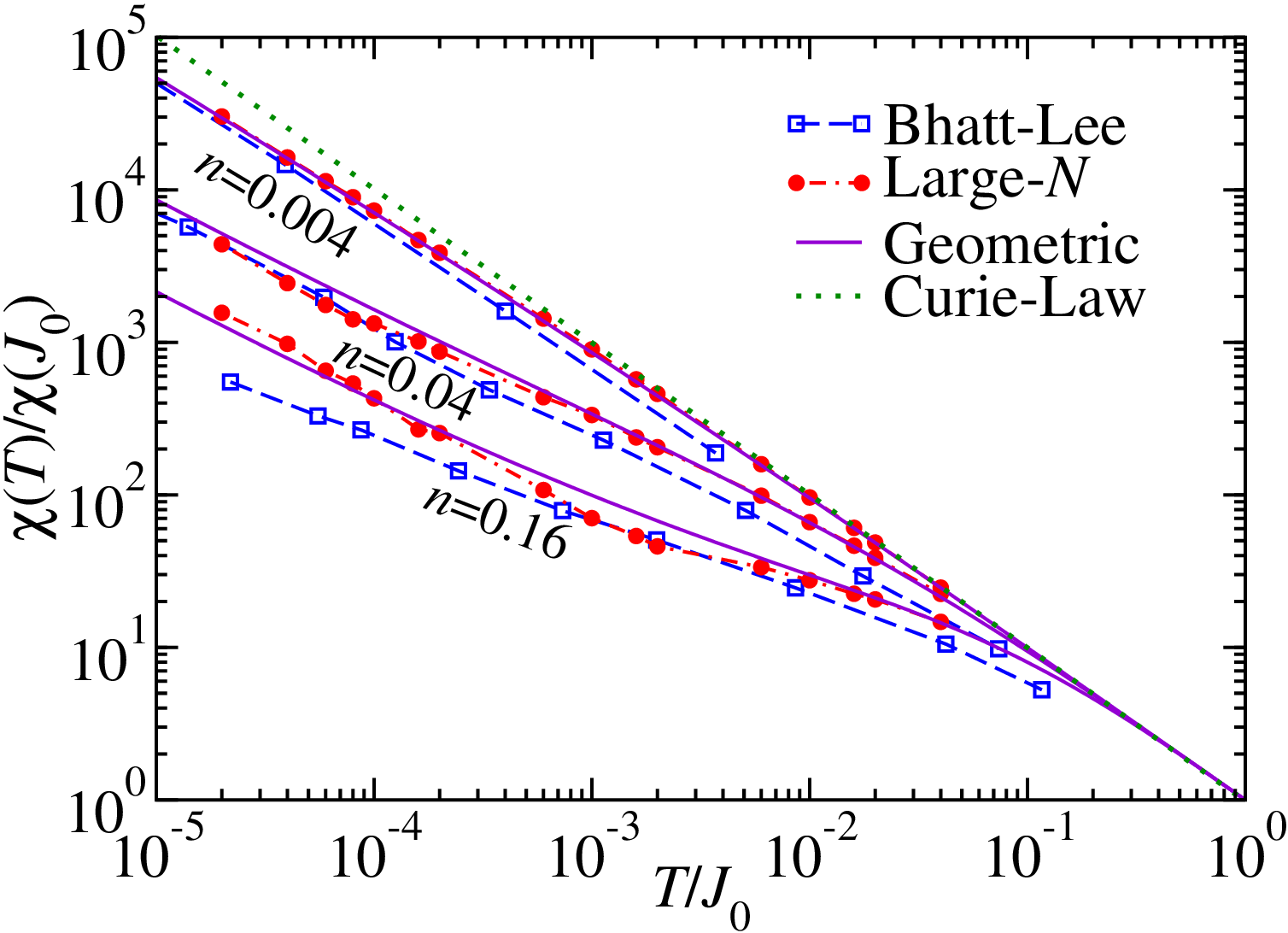}
\par\end{centering}

\caption{\label{fig1}(Colour online) Normalized magnetic susceptibility of
highly disordered 3D Heisenberg magnets evaluated with the Bhatt-Lee
method \cite{bhattlee82}, the large-$N$ self-consistent theory (for
systems with $N_{0}=512$ spins (see footnote \hyperref[foot2]{$^{\decimal{footnote}}$}), and the geometric decimation procedure at concentrations $n=\frac{4\pi}{3}\rho_{0}a^{3}=0.004$, $0.04$, and $0.16$.}
\end{figure}


\textit{Numerical large-N solution.} At any finite temperature, our
large-$N$ solution finds two types of spins: localized and VB spins.
The localized spins are those isolated from all other ones, \textit{i.e.},
$\Delta=0$ for all bonds connecting to them; their contribution to
the magnetic susceptibility is simply a Curie term $\chi_{c}(T)=\mu_{B}^{2}/k_{B}T$.
In contrast, each VB spin forms a singlet bond ($\Delta\neq0$) with
another spin; their contribution can be neglected at low temperatures.
The low-$T$ magnetic susceptibility is, therefore, well approximated
by 
\begin{equation}
\chi(T)=\rho(T)\chi_{c}(T),\label{chit}
\end{equation}
where $\rho(T)$ is the density of localized/free spins at temperature
$T$. Figure~\ref{fig1} shows the normalized magnetic susceptibility
$\chi(T)/\chi(J_{0})=J_{0}\rho(T)/T\rho_{0}$ at concentrations $n=\frac{4\pi}{3}\rho_{0}a^{3}=0.004$,
$0.04$, and $0.16$ (see footnote \hyperref[foot2]{$^{\decimal{footnote}}$}). The susceptibility diverges
at low temperatures, consistent with the SDRG results of BL \cite{bhattlee82}.
This divergence is usually fitted by a power law in experiments, but
we shall show later that it should be viewed as a logarithmic correction
to the Curie law. The higher the doping concentration, the larger
this correction since couplings among spins are stronger. At extremely
low concentrations, all spins are essentially free and the magnetic
susceptibility follows the Curie law. 


\textit{Geometric decimation procedure.} The large-\textit{$N$} ground
state at zero temperature of such a highly disordered Heisenberg system
is essentially a RS state, in which most spins form inert singlets
($\Delta=1$) with another spin and do \emph{not} correlate with any
other spin. To highlight this, we considered a simple four-spin cluster
with antiferromagnetic couplings $J_{ij}>0$, and $J_{23}\gg J_{ij}$
for all $(i,j)\neq(2,3)$. The large-$N$ calculation shows that for
$T>J_{23}$, all bonds are zero and all four spins are free. As we
lower the temperature to $J_{23}$, spins $\mathbf{S}_{2}$ and $\mathbf{S}_{3}$
start to form a VB singlet, $\Delta_{23}\neq0$, and no longer contribute
to $\chi(T)$. Further reducing the temperature to $J_{14}$, spins
$\mathbf{S}_{1}$ and $\mathbf{S}_{4}$ form another VB singlet. There
is \emph{no} resonance between the (2,3) and the (1,4) VB singlets.
In contrast to the Bhatt-Lee SDRG method, in which there appears a
renormalized coupling between SU(2) spins connected to a strong singlet
pair, this effect can be shown to be of order $1/N$ between SU($N$)
spins \cite{hoyos04}, and thus \emph{drops out} in the large-$N$
limit. While this simplification makes our large-$N$ model amenable
to closed form solution, we shall demonstrate that it hardly affects
the quantitative predictions of the model within the experimentally
relevant temperature range (as shown in fig.~\ref{fig1}).

\begin{figure}
\begin{centering}
\includegraphics[clip,width=1\columnwidth,keepaspectratio]{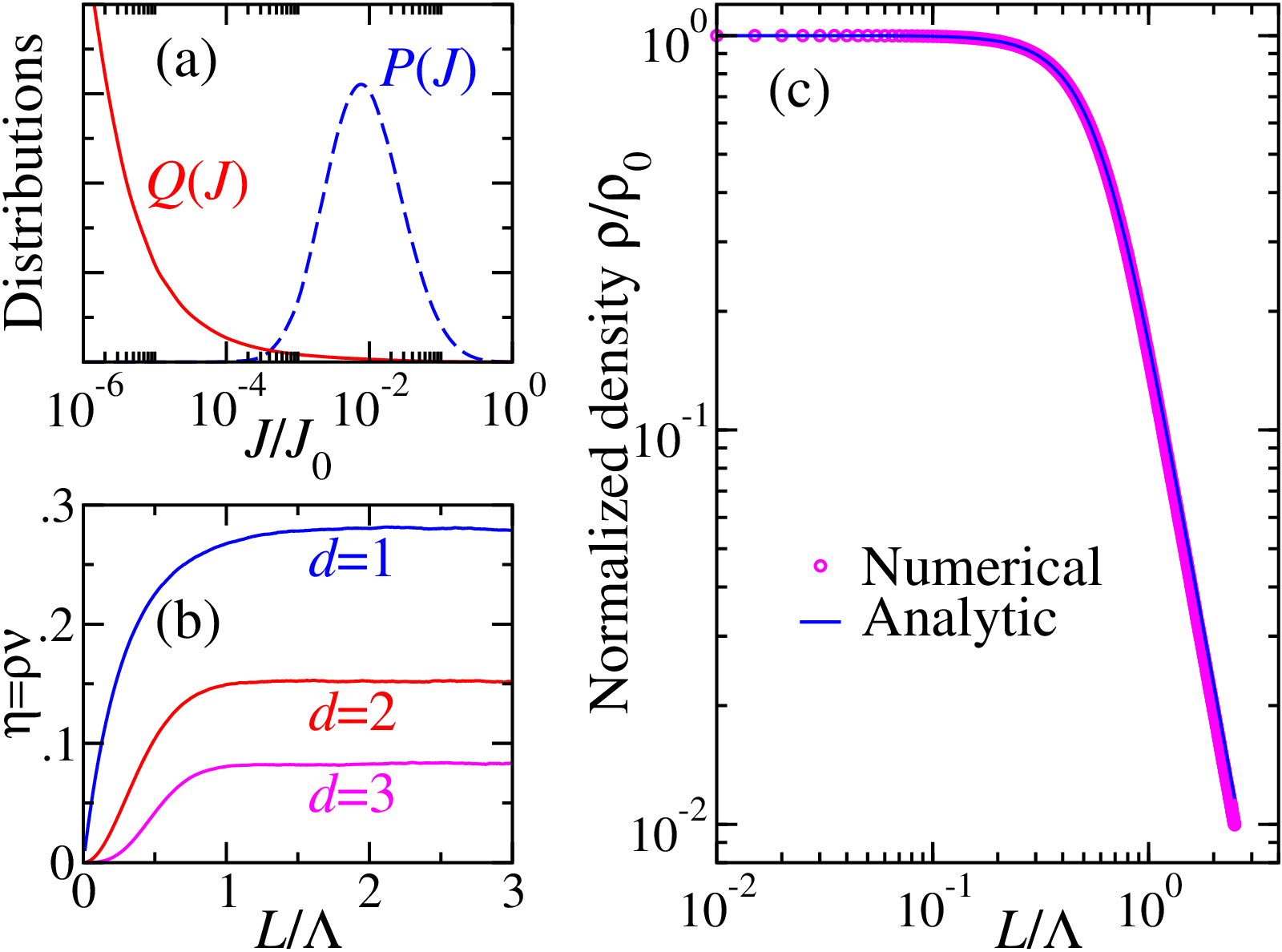}
\par\end{centering}

\caption{\label{fig2}(Colour online) (a) The distributions of the nearest
neighbor couplings $P(J)$ and of decimated couplings $Q(J)$ at concentration
$n=0.16$. (b) Numerical results for the packing fraction $\eta=\rho v$
as a function of decimation length $L$ for $d=1$, $2$, and $3$; $\Lambda=\rho_0^{-1/d}$.
(c) Comparison between the numerical and analytic (eq.~(\ref{rhov}))
results for the free spin density in the geometric decimation procedure
for $d=3$. Here, systems with $N_{0}=4\,096$ spins averaged over
$3\,000$ samples were used (see footnote \hyperref[foot2]{$^{\decimal{footnote}}$}).}

\end{figure}

This also allows us to state a very simple geometric decimation procedure.
We (i) search for the most strongly coupled spin pair, or equivalently,
the shortest one (see eq.~(\ref{tjs})),
(ii) remove it from the system by coupling the spins in an inert singlet,
and (iii) repeat steps (i) and (ii) until the desired energy (temperature)
scale is reached. We should emphasize that \textit{no} other renormalizations
are involved during this decimation procedure. The density of free
(undecimated) spins in eq.~(\ref{chit}) is then given by $\rho(T)=\rho_{0}\int_{0}^{T}Q(J)dJ$,
where $Q(J)$ is the distribution of the decimated couplings, shown
in fig.~\hyperref[fig2]{\ref{fig2}(a)} for $n=0.16$. The distribution
of nearest neighbor couplings, $P(J)$, is also plotted for comparison.
Note the dramatic difference between $P(J)$ and $Q(J)$ which stems
from the fact that, during the decimation procedure, longer-distance
nearest pairs are \emph{unavoidably} generated. Therefore, $Q(J)$
will always be singular yielding the divergence of $\chi(T)$ at low
temperatures. As depicted in fig.~\ref{fig1}, this simple geometric
decimation procedure captures the essential physics of the large-$N$
theory in describing the magnetic susceptibility of strongly disordered
Heisenberg spin systems. 

\textit{Analytic solution.} The geometric decimation procedure will
give us a long-sought analytic description \cite{analy1,analy2} of
the magnetic properties of insulating DS if one can keep track of
$\rho$ as a function of the energy scale $\Omega=\max\{J_{ij}\}$
(defined as the coupling to be decimated) or, equivalently, the length
scale $L=\min\{r_{ij}\}$ (the distance between the spins in the pair
to be decimated). Although the pair approximations \cite{analy1}
considerably simplify the calculations as compared to the SDRG and
numerical cluster calculations, they fail to yield an analytic expression
for $\rho$. On the other hand, the analytic formula proposed by Ponomarev
\textit{et al}. involves a tunable parameter \cite{analy2}. Here
we present an accurate analytic solution without any adjustable parameters
for a general $d$-dimensional system. 

Since we remove hierarchically the closest spin pair, we can imagine
each spin as a hard sphere of diameter $L$, which naturally incorporates
the constraint that no spin pair is closer than $L$ 
(see footnote \footnote{\label{foot-neglect} In the following, we neglect any other correlations
beyond those imposed by the hard-sphere constraint.}).
By removing the spheres that are touching each other, we continuously
increase $L$ until the next closest pair of spins touch each other.
The rate equation governing the density of free spins is given by
\begin{equation}
\mathrm{d}\rho=-2^{d}\rho^{2}g\mathrm{d}v,\mbox{ where }g(\rho)=(1-\alpha\rho v)(1-\rho v)^{-d}\label{rate}\end{equation}
 is the radial distribution function \cite{song89} of a hard-sphere
fluid: it gives the ratio of the density of particles at distance $r$ by the mean density, given that
there is a particle at the origin. Here, $\alpha$ is a constant which depends only on dimensionality
($\alpha=0$, $0.436$, and $0.5$ for $d=1$, $2$ and $3$, respectively)
\cite{song89}, and $v$ is the excluded volume of each hard sphere.
The negative sign comes from the fact that $\rho$ decreases as $L$
increases, and the decrease in $\rho$ is proportional to the density
of available spins $\rho$ times the probability that two spins (hard
spheres) touch each other, \textit{i.e.}, $2^{d}\rho g\mathrm{d}v$.
The $2^{d}$ factor converts the radius of the hard sphere (raised
to the power $d$) into its diameter. 

The solution of eq.~(\ref{rate}) can be reduced to a quadrature,
from which we can deduce that the packing fraction $\eta=\rho v$
increases monotonically with $L$, saturating at large length scales
at $\eta_{c}$ ($\simeq0.333$, $0.182$, and $0.0968$, respectively,
for $d=1$, $2$ and $3$). The results of a numerical solution of
the decimation procedure are shown in fig.~\hyperref[fig2]{\ref{fig2}(b)},
from which we obtain $\eta_{c}\simeq0.2810(5)$, $0.156(1)$, and
$0.082(2)$ for $d=1$, $2$ and $3$ (see footnote
\footnote{Small differences between the analytic and
the numerical values of $\eta_{c}$ reflect the higher order correlations
we have neglected (see footnote 
\addtocounter{footnote}{-1} \hyperref[foot-neglect]{$^{\decimal{footnote}}$}).}). Since
$\eta\ll1$ \textit{throughout} the decimation procedure, our hard
sphere liquid remains moderately correlated (away from the strong coupling
regime in the vicinity to close packing). This provides a dramatic
simplification, since we are now well justified in using the virial
expansion $g^{-1}\approx1-(d-\alpha)\rho v$ (this linearized expression
is exact \cite{song89} in $d=1$), and find a closed form solution
\begin{equation}
2^{d}\gamma\rho v=1-\left(\rho/\rho_{0}\right)^{\gamma},\mbox{ with }\gamma=1+\left(d-\alpha\right)/2^{d},\label{rhov}
\end{equation}
 which satisfies the initial condition $\rho=\rho_{0}$ at $v=0$.
The magnetic susceptibility in eq.~(\ref{chit}) is readily obtained
by relating temperature and $L$ via eq.~(\ref{tjs}), \textit{i.e.},
$2L=a\ln(J_{0}/T)$. In the $L,v\rightarrow\infty$ ($T\rightarrow0$)
limit, the density decays asymptotically as $\rho\sim v^{-1}\sim L^{-d}$.
Thus the magnetic susceptibility diverges at low temperatures according
to \begin{equation}
\chi(T)\sim\frac{J_{0}}{T\left[\ln\left(J_{0}/T\right)\right]^{d}},
\label{logcorr}
\end{equation}
 which can be viewed as a logarithmic correction to the Curie law
instead of the power law divergence usually fitted to experiments.
The free spin density $\rho$ extracted from eq.~(\ref{rhov}) is
plotted in fig.~\hyperref[fig2]{\ref{fig2}(c)} as a function of
$L$, in excellent agreement with the numerical result of the decimation
procedure. Therefore, eq.~(\ref{rhov}) provides an accurate analytic
solution, without any adjustable parameters, to the large-$N$ theory
of the insulating DS. 

\begin{figure}
\begin{centering}
\includegraphics[clip,width=0.9\columnwidth,keepaspectratio]{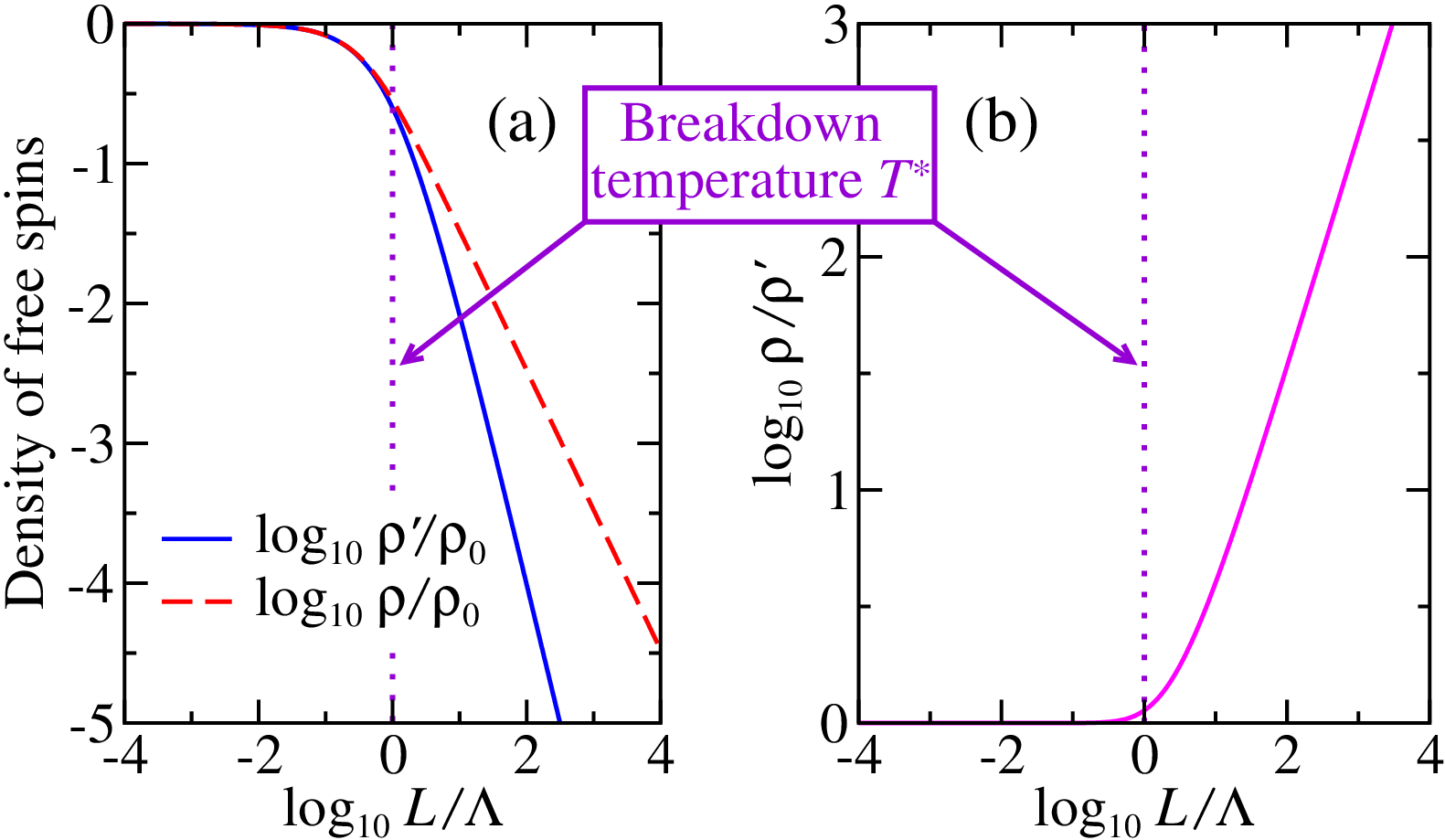}
\par\end{centering}

\caption{\label{fig3}(Colour online) Comparisons between the densities of
free (undecimated) spins as functions of the length scale $L$ obtained
by the SDRG ($\rho^{\prime}$) and the large-$N$ ($\rho$) methods,
i.e., the geometric solution, in $d=1$. The vertical dotted line highlights the 
breakdown temperature $T^*$ ($47\,\mathrm{mK}$ for $a\rho_{c}=a/\Lambda_c=0.25$ 
(see footnote 
\addtocounter{footnote}{0} \hyperref[foot-T]{$^{\decimal{footnote}}$})).}
\end{figure}

\textit{Comparison between SDRG and large-N.} It is now natural to
ask how reliable the large-$N$ theory is. To address this issue,
we compare the well-known RS solution of the 1D random Heisenberg
system obtained by the SDRG method \cite{fisher94,exactSDRGflow}
with the analytic solution, eq.~(\ref{rhov}), of the large-$N$ theory.
For randomly distributed spins, the length distribution of the nearest
neighbor bonds is a Poissonian $P(L)=\rho_{0}\exp(-\rho_{0}L)$, which
gives rise to a power-law initial coupling constant distribution 
\begin{equation}
P_{0}(J)=\theta(J)\theta(J_{0}-J)\frac{\rho_{0}a}{2J_{0}}\left(\frac{J_{0}}{J}\right)^{1-\rho_{0}a/2}.\label{pj0}
\end{equation}
 In this case, the SDRG flow can be followed exactly through all energy
scales, yielding \cite{exactSDRGflow}
\begin{equation}
\rho^{\prime}=\rho_{0}\left[1+\frac{\rho_{0}a}{2}\ln\left(J_{0}/\Omega\right)\right]^{-2}=\rho_{0}\left(1+\rho_{0}L\right)^{-2},
\label{rhop}
\end{equation}
 where the prime is added to distinguish this SDRG density from the
large-$N$ result in eq.~(\ref{rhov}). In the asymptotic $L\rightarrow\infty$
limit, $\rho^{\prime}\sim L^{-2}$, different from the $L^{-d}$ behavior
of the the large-$N$ theory as shown in fig.~\hyperref[fig3]{\ref{fig3}(a)}.
However, upon close inspection, the $L$ dependences of $\rho$ and
$\rho^{\prime}$ (see fig.~\hyperref[fig3]{\ref{fig3}(b)}) reveal
that the breakdown occurs only above a length scale $L^{\ast}=1/\rho_{0}=\Lambda$,
corresponding to a breakdown temperature 
\begin{equation}
T^{*}=J_{0}\exp(-2\Lambda/a)
\label{eq:breakdownT}
\end{equation}
below which the renormalized couplings become important in the SDRG
procedure. Above $T^{*}$, however, the SDRG theory can be reduced
to the simple geometric decimation procedure. The smaller the concentration
$\rho_{0}$, the lower $T^{*}$ is. Since $T^{*}$ concerns only the
energy scale at which the renormalized couplings become important,
eq.~(\ref{eq:breakdownT}) straightforwardly holds in higher dimensions.
Interestingly, this result implies that a large class of highly disordered systems
can be described by the random singlet picture above $T^{*}$ even
though their ground states are completely different (as they are in
Refs.~\cite{2D-dis-Heis}). For instance, $T^{*}\approx47\,\mathrm{mK}$
when $a\rho_{0}^{1/3}=a\rho_{c}^{1/3}=a/\Lambda_c=0.25$, assuming $J_{0}=140\,\mathrm{K}$
from Ref.~\cite{jvalue}\addtocounter{footnote}{1}\footnote{\label{foot-T} Corrections to $J_{ij}$ in eq.~(\ref{tjs}) may be important in order to compute the
precise value of $T^*$. If one naively inserts the factor of $(r_{ij}/a)^{5/2}$, one gets that 
$T^{*}=J_{0}(\Lambda/a)^{5/2} \exp(-2\Lambda/a)$ instead of eq.~(\ref{eq:breakdownT}), which increases $T^*$ by a factor of $32$ at
the critical concentration. However, note that $\Lambda/a = 4$ is not much greater than $1$. Thus, subleading corrections of 
order $(r_{ij}/a)^{2}$ \cite{andre81} become important. To our knowledge, they are not known at the present moment.}. Remarkably, the temperature window relevant
for experiments is above the breakdown temperature (left of the dotted line in fig.~\ref{fig3}), which also explains
the success of the BL theory. 

Finally, we would like to call attention to a caveat on
 eq.~(\ref{logcorr}). As shown in  fig.~\ref{fig3}, the experiments take place in
a temperature range above  $T^*$ in which both the SDRG and
the geometric decimation solutions coincide \emph{and} before their
asymptotic regimes have been reached. It is thus very clear
that either the numerical solution or the analytic one in
 eq.~(\ref{rhov}) compare well with experiments. In
 $d=1$ and above $T^*$, eq.~(\ref{rhov}) can be well approximated
by  eq.~(\ref{rhop}). Again, it is clear that the apparent power-law divergence
of the susceptibility seen in experiments should instead be
interpreted as a logarithmic correction to the Curie law.


\textit{Summary and outlook.} We have shown how a variational large-$N$
method provides a physically transparent and quantitatively accurate
description of inter-site spin correlations on the insulating side
of DS. In the presence of strong positional disorder, each localized
spin forms a VB singlet with a rather uniquely defined partner, allowing
for a closed-form solution of the problem in the large-$N$ limit. 

Even more importantly, this approach opens a very attractive avenue
to describe the behavior across the MIT. It is known that the large-$N$
RVB approach correctly describes the high density Fermi Liquid state~\cite{affleck-marston88}.
As we established that it also works in the opposite (insulating/Bhatt-Lee)
limit, then it will also provide a valid description of the transition
by examining the two-orbital $t$-$J$ model of eq.~(\ref{htj})
with finite inter-site hopping $t_{ij}$. Each $\tilde{f}$-spin now
has more than one choice: to still form a VB singlet with another
localized moment, or to undergo Kondo screening by conduction electrons.
Similarly as in the large-$N$ solution of the two-impurity Kondo
problem \cite{jones89prb} , we expect Kondo-screened sites to contribute
to the formation of a coherent Fermi liquid, while VB singlet pairs
to {}``drop out'' from the conduction sea and remain Mott localized.
Such gradual conversion of the correlated electron fluid into a localized
VB solid may provide a microscopic underpinning for the phenomenological
{}``two-fluid'' model \cite{paalanenetal88} - possibly the key
missing link for cracking the metal-insulator transition in doped
semiconductors. 

We thank Ravin Bhatt for useful discussions, and Richard Stratt for
very helpful comments about the hard sphere problem. This work was
supported by FAPESP grant 07/57630-5 (EM), CNPq grant 305227/2007-6
(EM), and by NSF grants DMR-0506953 (JAH) and DMR-0542026 (VD).

\end{document}